# Aiding Long-Term Investment Decisions with XGBoost Machine Learning Model

Ekaterina Zolotareva[1][0000-0002-1516-7378]

[1]Financial University under the Government of the Russian Federation,
38 Shcherbakovskaya St., Moscow 105187, Russia

`ELZolotareva@fa.ru`

**Abstract.** The ability to identify stock market trends has obvious advantages for investors. Buying stock on the upward trend (as well as selling it in case of downward movement) results in profit. Accordingly, the start and endpoints of the trend are the optimal points for entering and leaving the market. The research concentrates on recognizing stock market long-term upward and downward trends. The key results are obtained with the use of gradient boosting algorithms, XGBoost in particular. The raw data is represented by time series with basic stock market quotes with periods labelled by experts as «Trend» or «Flat». The features are then obtained via various data transformations, aiming to catch implicit factors resulting in the change of stock direction. Modelling is done in two stages: stage one aims to detect endpoints of tendencies (i.e. "sliding windows"), stage two recognizes the tendency itself inside the window. The research addresses such issues as imbalanced datasets and contradicting labels, as well as the need for specific quality metrics to keep up with practical applicability. The model can be used to design an investment strategy though further research in feature engineering and fine calibration is required. This paper is the full text of the research, presented at the 20th International Conference on Artificial Intelligence and Soft Computing Web System (ICAISC 2021).

## 1   Introduction

The ability to identify stock market trends has obvious advantages for investors. Buying stock on an upward trend (as well as selling it in case of downward movement) results in profit, which makes predicting stock markets to be a highly attractive topic both for investors and researchers.  Sure enough, since the 1970s various methodologies have developed: fundamental analysis, technical analysis, time series econometrics, fuzzy logic, etc. are used to detect trends.  Despite the long history, this field, as stated in [1], is still a promising area of research mostly because of the arising opportunities of artificial intelligence. With the rise of machine learning in the early 2010s, researchers started to take interest in applying computer science to financial market problems [1–3]. Machine learning can be thought of as an extension or even an alternative to traditional statistical methods. Instead of using specific parametric models to explain dependencies, machine learning aims to find hidden or poorly structured dependencies in data by learning from a vast number of examples.

## 2      Literature overview

The survey [1] provides a very thorough literature overview, which gives the historical outline of financial time series prediction, as well as the review of the main path in the literature on financial market prediction using machine learning. The authors have studied and classified 57 articles, covering the period from 1991 to 2017. In general, there are three main classes of models - artificial neural networks (ANNs), support vector machines (SVM/SVRs) and various decision tree ensembles (e.g., random forests). As of 2017, the hegemony of ANNs and SVM/SVRs has been observed - the articles based on these models accounted for 86% of articles researched in [1]. Decision trees and Random forests as the main forecasting technique were used only in 7 out of 57 selected articles. This may be explained by the relatively shorter history of scientific research in the case of decision tree ensembles: while many papers on ANNs and SVM/SVRs date back to the early 2000s [1], the papers on applying random forests to stock market forecasting appear mostly in 2010-2015[4–9].

To get a better picture of the current state of research in applying decision trees to stock market forecasting, we provide a brief overview of relevant scientific papers, published in the period between 2018 and early 2021. The classification of studies used in this paper resembles the one suggested in [1] with minor alterations. The search in the Scopus database was performed on 10.10.2020 and was tuned in order to find the most relevant articles from Q1/Q2 journals. In total, twenty papers, containing original models, from four journals were reviewed (**Fig. 1**).

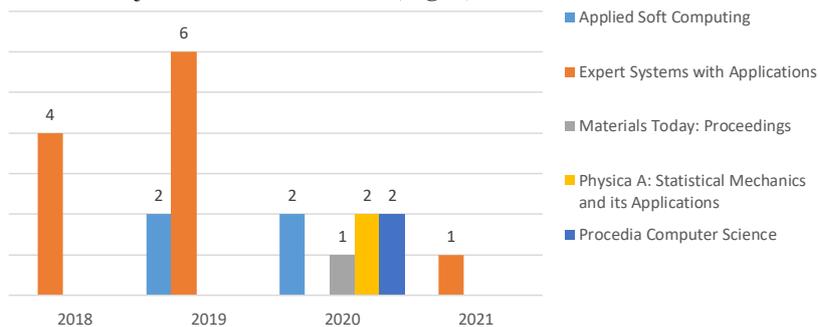

**Fig. 1.** The reviewed articles grouped by journal name and year.

Same as in previous years, artificial neural networks (ANN) prevail as the method of prediction (see **Fig. 2**) being used as the main algorithm or at least for comparison reasons (like in [10–13]). Only five groups of researchers [14–18] out of twenty do not use ANNs at all. In nine studies [19–26] ANNs were exploited as the only main algorithm. Decision tree algorithms were solely used as the main method only in three studies [13, 14, 27], as well as support vector machines/regressions [10, 17, 18]. Among other models are logistic regression [13, 14, 16], linear regression [28, 29], extreme learning machine and ARMA [12].

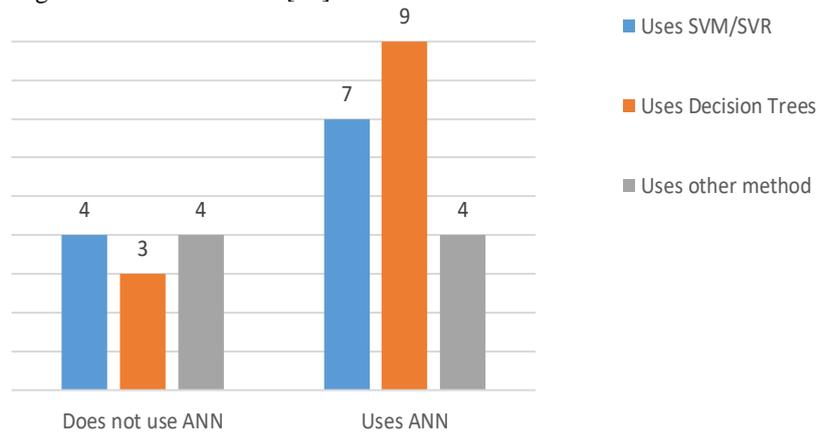

**Fig. 2.** Machine learning methods implemented in the reviewed articles

Besides, several researchers have exhibited model fusion – an approach discussed in [3] as a process of combining various factors (e.g. models) that can improve the performance and provide useful results. For example, in [13] the authors cascaded logistic regression onto gradient boosted decision trees for forecasting the direction of the stock market. In [30] the researchers combined ANN, SVR, random forest and boosted decision trees to predict short-term stock prices. The stacking methodology was demonstrated in [31] by fusing the outputs of four types of tree ensemble models and four types of deep learning algorithms for stock index forecasting. Model fusion appears also in combining machine learning predictions for the aims of portfolio management [17, 27].

As we can see, neural networks tend to supersede SVM/SVR (which had hegemony according to the previous literature review [1]), while the role of decision tree algorithms (decision trees/random forests/gradient boosting) remains stable, though still small. Therefore the opportunities of these types of models are not yet fully explored, especially considering the velocity of progress in the artificial intelligence field.

It should be noted that despite the common main method different researches have sufficient variations in model structure, problem formalization, and features and labels accordingly. Models also may be applied to different markets, assets and prediction horizons. A certain variation also exists in the selection of performance measures which should ensure the comparability of models.

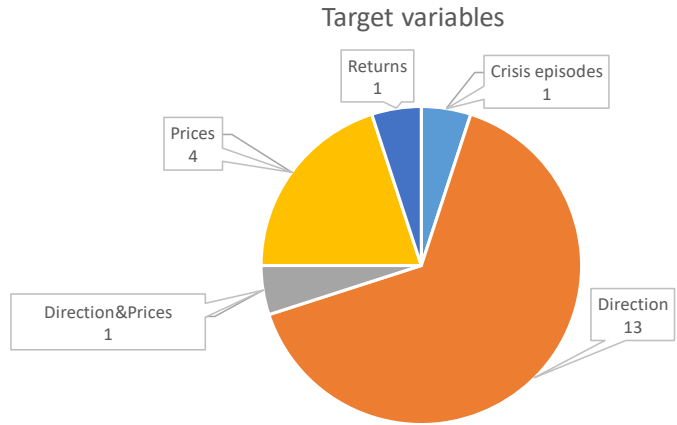

**Fig. 3.** The difference in target variables for the articles reviewed

As shown in **Fig. 3**, most researches are concentrated on predicting the direction of the stock market thereby solving a classification problem. Fewer predict prices [22, 23, 28, 30] or returns [11] by solving a regression problem. One study suggests models for both directions and prices [12] predictions introducing a hybrid variational mode decomposition and evolutionary robust kernel extreme learning machine. In all the cases the ground truth variable is extracted from the historical price series. The type of model (classification or regression) also predefines the choice of quality metrics: accuracy, AUC and F1-Score – for classification tasks and MAE, RMSE, MAPE – for regression. Also, the returns of simulated trading can be compared [18, 24].

Among the "direction type" models, the absolute majority aim to predict the next day price movement, which can be considered a short-term objective. Only in four studies the time horizon varies from one week [20, 25] to one month [31] or even several months [16]. Actually, as shown in **Fig. 4**, the focus on daily basis predictions prevails within the reviewed articles for all types of target variables, except for the study [19] - but this is due to the specifics of the research problem - detecting crises episodes.

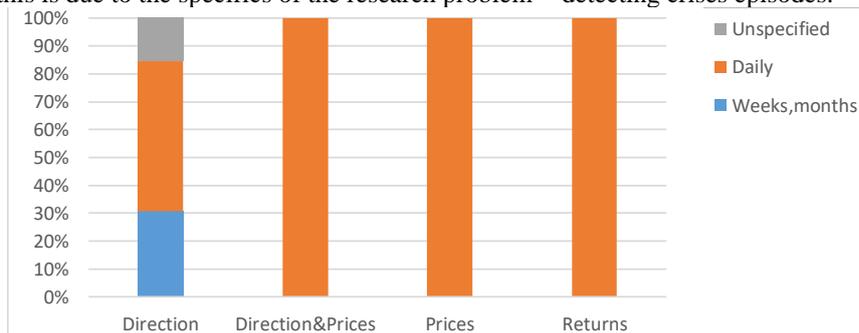

**Fig. 4.** The difference in time horizons of predictions for different target variables

Another important difference between the suggested models is the choice of features (see **Fig. 5**). Though most of the researches used market data (e.g. prices, volumes and the values derived from them - technical analysis indicators, correlations, volatilities and returns) as input variables, two studies used – and with success – only news [14] and investor sentiment [18] for the prediction of price direction. Text features were also used in [25] and [30]– along with technical analysis indicators (TA). Fundamental features (e.g. oil prices) were only found in two studies [24, 31] combined with TA and raw prices. If compared to the previous literature review, the proportion of studies that exploit fundamental variables is relatively smaller -according to [1] they were used in 26% of articles between 1991-2017. But in recent years we again see the implementation of the information fusion approach [3] in the form of feature fusion.

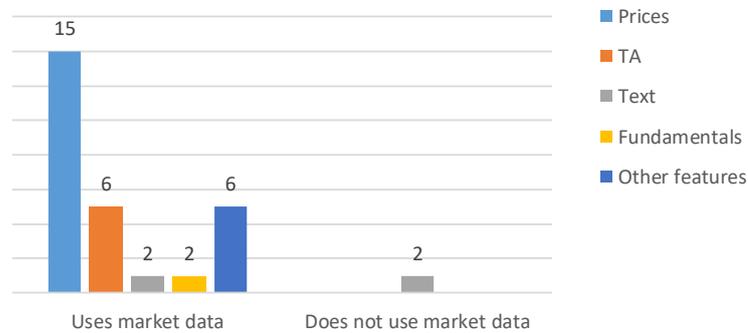

**Fig. 5.** The feature fusion in the reviewed articles

The geography of the research (**Fig. 6**), same as in the previous years, is concentrated on the prediction of the USA, Europe and China stock markets, or all of them together. This can be explained by the availability of such data, which is useful for the comparison of models, and by the boost in academic research in China, noted in[1]. Nevertheless in three studies emerging markets were explored: Malaysia [29], India [12] and Brazil [17].

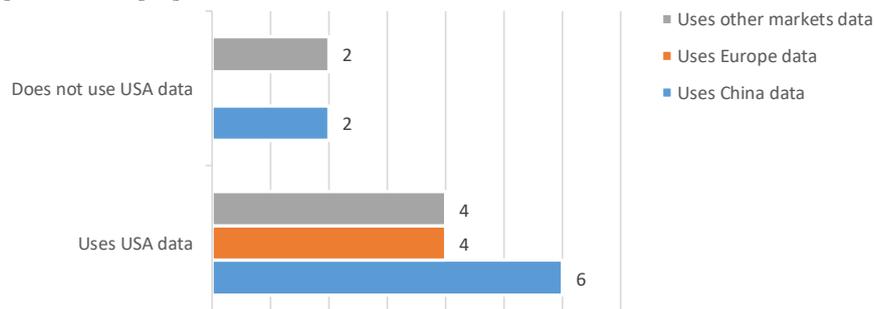

**Fig. 6.** The geography of data used for research

In this paper, we apply a gradient boosting algorithm, specifically XGBoost, to the problem of finding the start and endpoints of trends. Gradient boosting is a relatively new algorithm, suggested by [32] in 2016. It is a decision tree ensemble method that was designed to fix some of the faults of random forests. More specific information on the algorithm will be given in section 7.4. Gradient boosting has several technical realizations, the most known being XGBoost, which is extensively used by machine learning practitioners to create state of art data science solutions [33].

Unlike the majority of other researchers, we aim to explore long-term trends, which last several months, not days. The start and endpoints of such trends accordingly are the optimal points for entering and leaving the market. Despite the various technical analysis algorithms and econometrics studies based solely on stock data, some market experts still argue that traders are able to see opportunities of making money (i.e. detecting trends or turning points) that cannot be formally expressed. Thus, using computer science algorithms to learn from successful traders' decisions (and not only stock data) is likely to improve financial market models. The key distinction of our model is that its ground truth vector is fully based on expert opinion data, provided by one major Russian investment company. Unlike other researchers, we do not use a mathematical formula to define a trend, instead, it is defined by an expect as a potentially profitable (or unprofitable) pattern in price dynamics. The major difficulty of this approach is that it is exposed to subjective judgements of the experts. On the other hand, if the experts are successful traders in a certain investment company, it gives the employer the chance to 'digitalize' their exceptional skills and obtain a machine learning algorithm no one else on the market can employ.

The training is performed on historical S&P stock data with additional feature engineering. The model requires only raw historical price data as input. From one point of view, it can be considered a limitation, since we ignore fundamental factors and news feed. On the other hand, it makes the model unpretentious in production, since stock data is easily obtainable and can be downloaded into the company's informational systems or directly fed into the model via API.

Technically we are solving a classification problem (it is a "direction type" research), but the standard quality metrics (accuracy, AUC and F1-Score) turn out to be inapplicable because of imbalanced datasets and contradicting labels. For these reasons, the returns of simulated trading are used as the main performance indicator. The returns of simulated trading are used as the main performance indicator.

## 3 Problem formulation

The research was conducted on behalf of one major Russian investment company (the Company). The Company experts have labelled historical S&P stock data, that is, they marked certain consequent periods as "Trend"," Flat" or N/A in a specially designed software with a graphical interface (see **Fig. 7**).

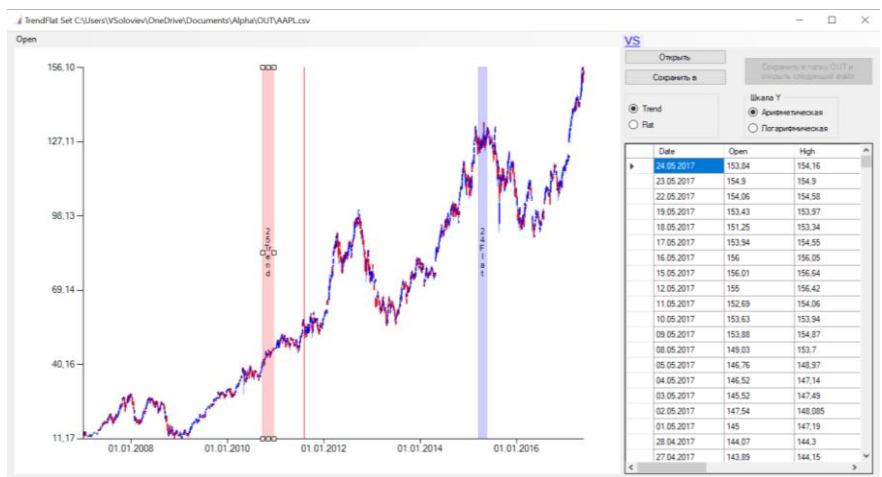

**Fig. 7.** GUI interface for data labelling

Approximately 90% of identified trends last between 40 to 600 business days, which accounts for middle- or long-term tendencies. Initially, the task was to train the model to identify the trend itself (no matter the direction) with the minimum lag from its start, for example, to recognize a 200-day-long trend 20 days after it started. Later, though, it turned out that it is also necessary to distinguish between downward or upward trends in order to calculate and compare the financial results of different strategies. The model should be independent of any specific stock, market, or time period, after it is properly calibrated it should be equally good for any asset and time. Another important issue is that by learning from historical patterns we aim to identify the current market situation (answer the question: what long-term tendency takes place today?) and we must always bear in mind that future data is unavailable. Breaking this condition will make the modelling results irrelevant, though minor time lag (within a couple of weeks) is quite acceptable.

## 4 Data overview

The dataset to explore consists of two sources described in **Table 1** below.

**Table 1.** The dataset to explore.

|  | Source I | Source II |
|---|---|---|
| Date received | June 2017 | October 2017 |
| Total number of files | 700*[1] | 3162 |

---

[1] *excluding 16 defect files containing 43 918 records. Due to technical errors these files either contained quotes from a different marketplace or had a wrong stockname. Mixing this data with the remaining datasets would result in having different quotes for the same date and stockname.

| | Source I | Source II |
|---|---|---|
| List of experts | 7 experts – **A**,**B**,C,**D**, E,F,**G** | 6 experts - **A**,**B**,**D**,**G**,H,I |
| Total number of records | 1 706 441* | 7 474 271 |
| Excluding duplicates | | |
| Total number of stocks | 106 | 599 |
| Time period | From 2005-01-28 to 2017-05-24 | From 2007-08-08 to 2017-09-13 |
| Data overview | The datasets have an intersection in the time period - dates from 2007-08-08 to 2017-05-24, but they have only one intersection in the list of stocks – only stock named "MS" is present in both datasets. Only 4 experts labelled both Source I and Source II data, but the second dataset is considered "cleaner" since the experts were more motivated to label data responsibly. | |

The total number of records in both datasets amounts to 9 180 712 pieces packed in 3162 files. Each file contains on average around 2600 daily quotes for a certain period and stockname, labelled by a certain expert. The description of the main fields is given below (**Table 2**):

**Table 2.** The description of the main fields.

| Field name | Description | Notes |
|---|---|---|
| Date | Date | These fields form a "data point". They are independent of experts. The values in this field for a given date and given stockname must be unique. |
| Open | Day opening price | |
| High | Day maximum price | |
| Low | Day minimum price | |
| Close | Day closing price | |
| Volume | Trading volume | |
| Stockname | Stockname | |
| IDselect | The ordinal number of the window – a time period within which the tendency remains constant. In each file the enumeration of windows is reset. | These fields are expert-dependent. Each expert would label a certain data point only once, but different experts can label the same data points and their expertise does not necessarily coincide. This results in data inconsistency – the issue which would be addressed later in the paper. |
| Type | Type of tendency associated with a certain window. Can take values "Trend", "Flat" or "N/A" (upon convention with the Company N/A should be treated as "Flat" and not as a separate category) | |
| Username | Expert name. For the purpose of this paper, the experts are addressed as A, B, C, etc. | |

## 5  Model outline

Modelling is done in two stages: stage one aims to detect endpoints of tendencies ("change points", or "turning points"), stage two recognizes the tendency itself inside the window.

The performance of stage one is provided by the model which will be referred to as "ChangePoints". For each data point it returns either a value "1" ("The change of tendency occurred" or "A new window has started") or 0 ("No changepoint"). Due to various reasons, discussed later, currently, the ChangePoints predictions are subject to both false positive (mainly) and false negative errors. This is why it can't be used alone and should be backed by the stage two model - "TrendOrFlat". TrendOrFlat is launched when the possible start of a new tendency is detected, i.e. "Change points" returns "1". It is important to note that once the changepoint signal occurs, we come to recognize the starting point of the new tendency, but we do not know how long it will last or when the endpoint will occur. To identify the tendency inside the new window TrendOrFlat model initially analyzes the first few, say 6, days of it, then first 7 days, then 8, etc., returning the values "1" ("Upward trend"), "-1" ("Downward trend") or "0" ("No trend/Flat") for each period. Shortly after the start, TrendOrFlat is more likely to produce incorrect predictions, but as the window widens, the tendency identification should become more and more accurate. The process continues up until a new positive signal from ChangePoints occurs, indicating the start of a new window for which the routine repeats. The final results of modelling are determined upon TrendOrFlat output. The whole process, run for a set of instruments on the chosen time period, will be referred to as 'pipeline'.

## 6  Train and test sample

In order to evaluate the generalizing power of the model we traditionally divide the dataset into train and test samples. These samples should be independent, otherwise, the quality metrics would be misleadingly inflated. We considered three possible options for train/test split with the more or less standard proportion of 70 to 30:

1. Standard random split for records turned out to be totally inapplicable in our case. As mentioned before, different experts could label the same data points. Sometimes their expertise would match. This results in 6 199 781 (67,5%) duplicate records for the ChangePoints model. They would be present in both train and test samples and thus break the condition of independence. We could of course manually exclude duplicates from the ChangePoints dataset. But since ChangePoints and TrendOrFlat use different sets of features, it would still be complicated to exclude intersection between train and test samples for these models (an element from the test sample for ChangePoints might appear in TrendOrFlat train sample and vice versa);
2. Split by source. The proportion between the number of records in Source I and Source II datasets is roughly 80(Source II) to 20 (Source II) which makes a natural

10split between train and test. Except for one stockname, which hardly accounts for 0,2% of records, the datasets do not have intersections in labelled datapoints. However, the correlation arises from the common time period: from 2007-08-08 to 2017-05-24. Since the quotes are taken from the same marketplace, even if the stocknames are different, they still follow global market tendencies present in a certain time period. This again will break the condition of independence, though less obviously.
3. Split by date. To ensure the independence of data, our final choice is to use 70% of older data as a train set and the remaining 30% as a test set. The split date is October 14$^{th}$, 2014 if both Source I and Source II are analyzed and November 6$^{th}$, 2014 if only Source II is used.

## 7    ChangePoints model

We start with the ChangePoints model and its features. It so happened that this model is subject to various complicated issues, while TrendOrFlat works fairly smoothly.

First of all, we need to deal with the target variable – we shall call it "NewTrigger". It should take the value "1" the day the tendency changes and remain "0" otherwise. The original dataset contains the "Type" feature and the "IdSelect" feature. The change in either of these values can be thought of as a changepoint. The natural and easier choice would be to extract the "NewTrigger" variable from the change in the "Type" value. However, this leads to missing a number of crucial changepoints, where the trend changes its direction. For example, if a downward trend is closely followed by an upward trend, both these tendencies will be marked as "Trend", while, obviously, they are absolutely different. Luckily such few but important cases were marked by different "IdSelect" values, which made it possible to determine "NewTrigger" without loss of accuracy.

As for the input features, they should comply with the task conditions: the model should be independent of any specific stock, market, or time period and can have only a minor time lag. That means we can't use date, stockname and absolute price or volume values as input variables (since they are stock-specific), neither can we scale by such values if they are not available at a given date – say, year average will be unknown until the end of the year. On the other hand, to see the change in tendency, we need to compare data before and after the changepoint, so some time lag is inevitable. We opted for 5 business days before and after the current date, since it roughly corresponds to one business week. The use of older data (6 days before today, etc) is also possible but will result in a number of additional variables. The use of future data determines the lag of the model and its increase is undesirable. The list of ChangePoints features is presented below (see **Table 3**), totaling 22 input variables plus 1 target.

**Table 3.** ChangePoints features

| Features | Description | |
|---|---|---|
| | Raw | Logarithmic |
| Close-1, Close-2,… Close-5 | Ratios of the 5 previous closing prices (today minus 1, minus 2 and so on) to the current closing price. | The natural logarithms of the corresponding ratios, or the difference between the logarithms of numerator and denominator |
| Close1, Close2,… Close5 | Ratios of the 5 future closing prices (today plus 1, plus 2 and so on) to the current closing price. | |
| Volume-1, Volume-2,…Volume-5 | Ratios of the 5 previous trading volumes (today minus 1, minus 2 and so on) to the current trading volume. | |
| Volume1, Volume2,…Volume5 | Ratios of the 5 future trading volumes (today plus 1, plus 2 and so on) to the current trading volume. | |
| High | The ratio of today's maximum price to the closing price | |
| Low | The ratio of today's minimum price to the closing price | |
| NewTrigger | Target variable, indicating whether the change in tendency has occurred today or not. | |

First, the raw features were used for modelling. Later it appeared more appropriate to use natural logarithms instead since they are less subject to the spreads in absolute values. Furthermore, it turned out that experts were using logarithmic price scales when labelling the data. However, the quality of the model didn't sufficiently improve just after introducing logarithmic variables. This fact has at least two explanations: first - the spread in absolute values within 2 business weeks is not that terrible to be improved by logarithms, second – other issues have a greater impact on the quality of the model.

### 7.1 Contradicting labels issue

It was previously mentioned, that when several experts label the same datapoints, their expertise does not necessarily coincide (see **Fig. 8**). There are three major reasons for that:

1. They are different traders. One would see a long upward trend, the other would see two middle-term trends interrupted by a short downward or flat period. Where the first sees no changepoints, the other would see two.
2. They may have misunderstood the task or completed it without due zeal.
3. Since the labelling was done in a GUI by mouse-clicking, technical blots might have occurred - one might fail to select the precise datapoint because of improper screen settings or poor eyesight.

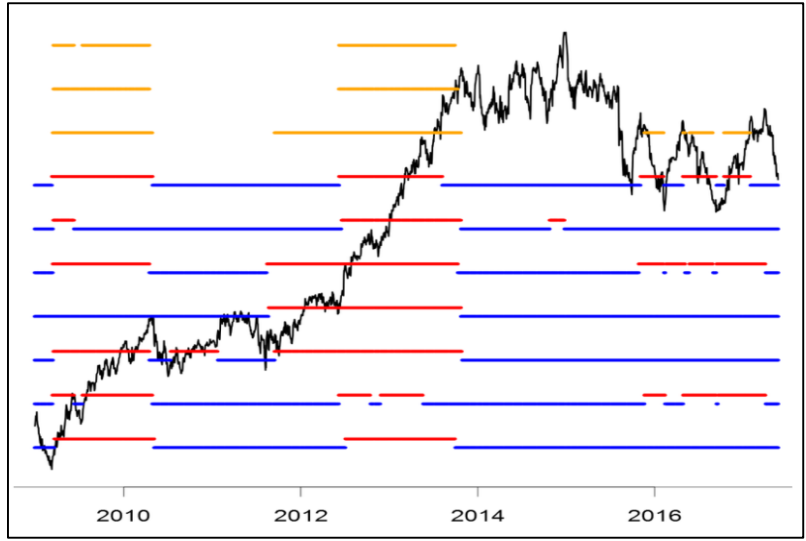

**Fig. 8.** Contradicting labels. Blue lines illustrate "Flat", red lines illustrate "Trend", orange illustrate regions where experts' opinions coincide (maybe partly).

Altogether this results in several thousands of records with identical input features but different target values. Literally <u>every</u> positive "NewTrigger" record has a negative contradict. To suppress this issue, the following strategies have been introduced:

1. Averaging the experts' opinions or voting. We convert the "Type" labels into numbers ("1" for the upward trend, "-1" for the downward trend and "0" for "No trend/Flat"), average them and round up. Suppose for a certain data point 3 out 6 experts (>=50%) say "1", and the others say "0, then the rounded average is "1" – upward trend. The pitfall of this approach is that it is majority-oriented and assumes all the experts are equally skilled or diligent.
2. Triggers correction. This alternation to data targets the technical blot issue. It forcibly pulls the start of the upward trend to the local (+/-5 days) minimum, and the start of the downward trend to the local maximum. Applying these corrections decreased the number of contradictions by 8- 18% depending on the dataset.
3. Excluding irrelevant experts. The best results of the modelling were achieved after leaving only experts D and G. These experts were the main stakeholders of the research, with greater experience and motivation.
4. Ignoring the contradictions. Having several opinions is not always a flaw. We may leave the contradicting records, assuming their share is small, and let the machine learn from these contradictions. The major pitfall of this approach is that traditional classification quality metrics (Accuracy, AUC, Precision, Recall, F-Score) will become irrelevant since there is no "ground truth" anymore[2]. The classification output is often presented in probabilities, the threshold of 0.5 is used by default to decide between classes. The contradictions are likely to "spoil" the classifier, but we can use custom threshold levels to control the level of confidence.

## 7.2 Imbalanced dataset issue

From a formal point of view, the ChangePoints model is a binary classification model. Basically, by training a model we mean finding the optimal set of parameters that will minimize the specific loss function. The loss function measures the discrepancy between the values of the target variable ("ground truth") and predictions, calculated for the whole sample. In a perfect situation, the proportion between classes should be close to 1:1, otherwise, the observations of a minority class would be "suppressed" by the majority. In other words, the mistakes on minority class observations would hardly influence the loss function. In highly imbalanced datasets this might lead to the total ignorance of the minority class, while this class can be especially important for the researcher. The traditional classification quality metrics, on the contrary, would perform quite well, unless you drill down into the contingency matrix or evaluate the performance on the minority and majority classes separately. Unfortunately, our dataset is highly imbalanced: records where "NewTrigger" equals "1" (positives) are the minority class. That is natural because we seek changepoints of middle-and long-term tendencies which would happen only once in several hundred business days.

Below are the figures for the list of selected ChangePoints models with different dataset options (see **Table 4**). Duplicate rows were excluded. As we can see, while averaging excludes contradictions, it makes the imbalance even more prominent.

**Table 4.** ChangePoints models with different dataset options. The figures are given for train sets only. The split date is October 14[th,] 2014 for Source I + II option November 6[th,] 2014 for Source II.

| Model Id | Logarithmic | Averaging | Trigger correction | Source (split by date) | Experts | Balance | Contradictions(% out of positives) |
|---|---|---|---|---|---|---|---|
| ChP20 | False | True | False | Source II | All | 357:1 | 0 |
| ChP22 | False | False | False | Source II | All | 76:1 | 12 443/ 100% |
| ChP26 | True | False | True | Source I+II | All | 78:1 | 14 987/ 99,5% |
| ChP27 | True | False | True | Source II | All | 98:1 | 10 163/ 99,8% |
| ChP28 | True | N/A | True | Source II | G | 290:1 | 0 |
| ChP29, 30 | True | False | True | Source I+II | D and G | 154:1 | 6 184/ 84% |
| ChP31 | True | True | True | Source I+II | D and G | 330:1 | 0 |

To contend with the imbalanced dataset issue, one can use special oversampling or undersampling techniques. They create a synthetic dataset, balancing classes by either boosting the minority or sampling from the majority. These synthetic datasets are then used to train the model. Another option is to keep balance by correcting the algorithm performance directly depending on whether we deal with the item from the majority or the minority class. No alterations to the initial dataset are needed in this case.

### 7.3 Quality evaluation issue

As we can see, there are at least two important groundings against traditional classification quality metrics (Accuracy, AUC, Precision, Recall, F-Score): they can be misleading both due to "ground truth" contradictions and highly imbalanced dataset. But we will find another reason to consider them irrelevant if we remember the time series issue. In a common classification model, the records are absolutely independent and if we shift the prediction of "1" to the neighboring record, we'll have a completely different result. But with the ChangePoints model, shifting one day back or forward will result in only a very minor change in terms of profit. This brings us to the conclusion that the most relevant quality metrics for such kind of models are those profit-related.

### 7.4 Changepoints XGBoost realization

There are a number of classification algorithms – e.g. logistic regression, Bayesian classifier, SVM, neural networks, decision trees and their ensembles. Out of all these diversities, it is gradient boosted ensembles that account for best modeling results during the last couple of years. The modelling in this research was completed with XGBoost (version 0.7.post3, Python 3.6), one of the technical realizations of gradient boosting [33]. The key idea of using decision tree ensembles instead of decision trees themselves is to address the problem of overfitting: while being able to perfectly describe the known data, single trees may fail to generalize dependencies. Technically this means that the prediction error will have a relatively low bias, but quite a big variance. It can be shown, though, that ensembles (or compositions) of trees have the same low error bias as single trees, but smaller variance. Moreover, the variance of the error is the less the lower the correlation between the algorithms. Random forests methodology suggests the way to build decision trees in such a way to make their predictions almost independent, which is achieved by training each algorithm on a different subset of initial data. Specifically, it employs bootstrapping for samples and random subsets for features. To keep low error bias, the trees need to be deep, but their number must be enough to provide low variance. These requirements, obviously, result in resource intensity, especially for large datasets. Gradient boosting suggests a different approach for ensembling single algorithms: the trees are added recurrently, and each new tree is designed in a way to correct the error of the composition of previously added trees [32].

Another advantage of the XGBoost algorithm is that it allows controlling the imbalanced dataset issue by directly setting the hyperparameter "scale_pos_weight" to the proportion between the negative (majority) and positive (minority) classes in the dataset ("balance"), ensuring the parity between the classes while training the model [34]. The best hyperparameters were searched by grid search/randomized grid search procedures, but after several iterations, it became obvious that it is mainly 4 hyperparameters that matter and even they do not alter the result crucially. Partly it is due to the label inconsistency issue, and partly - due to the highly imbalanced dataset, and of course – the feature choice might be improper. Besides we should remember that the final quality metric (profit) can be evaluated only after the whole pipeline is run and we can't yet formally assess the contribution of ChangePoints or TrendOrFlat to the final result. The ChangePoints traditional quality metrics however can be seen as a rough indicator of the overall model performance. They can be used to choose between concurrent ChargePoints modifications and fair enough we may assume that models with higher AUC or F-Score metrics are likely to give better results in terms of profit.

### 7.5 Selecting model hyperparameters

There are two sets of hyperparameters to consider while training the ChangePoints model.

The first set of hyperparameters concerns the data we use to train the model. As mentioned before, we may opt for logarithmic data instead of raw, choose to average experts' opinions or not, use trigger corrections or deal without them, as well as select certain sources (use both sources or only the "cleaner" second one) and experts. Varying these options results in different datasets (see **Table 4**).

The second set of hyperparameters are the XGBoost hyperparameters themselves currently totaling a minimum of 20 items. Some of them are purely technical and influence the speed and capacity of the algorithm, some can be set to default values, while others require a more attentive approach. To choose the best combination one can implement full grid search or randomized grid search procedures, though this might turn out to be very time-consuming. During this research, one full grid search and one enlightened randomized grid search procedures were conducted for the models with ids "ChP30" and "ChP31" accordingly.

The scope of grid search for "ChP30" covered the following parameter combinations can be found in **Table 5** (others are set to default).

**Table 5.** GridSearch options for "ChP30"

| Selected Model Hyperparameters | Grid | Description |
|---|---|---|
| n_estimators | 500,100 (default 100) | Number of boosted trees to fit. |
| max_depth | 3,7,10,15 (default 3) | Maximum tree depth for base learners |
| reg_lambda | 1,1.5, 2,5,10,0 (default 1)[2] | L2 regularization term on weights |
| scale_pos_weight | Balance =154 (default 1) | Balancing of positive and negative weights |
| learning_rate | 0.1 (default 0.1) | Boosting learning rate |
| reg_alfa | 0 (default 0) | L1 regularization term on weights |

---

[2] Due to the unexpected kernel shutdown only the options reg_lambda =5 and 10 were tried for the combination n_estimators=100 and max_depth=15. This however did not influence the conclusions.

| Selected Model Hyperparameters | Grid | Description |
|---|---|---|
| subsample | 0.8 (default 1) | Subsample ratio of the training instance |
| seed /random_state | 42 | Random number seed. Fixing the seed ensures we can reproduce the run |
| n_jobs/ nthread | -1 | The number of parallel threads used to run XGBoost. "-1" means using all possible threads. |

The scoring criterion was set to 'f1_macro' (the mean of the binary F1-score metrics, giving equal weight to each class) to highlight the performance of the minority class. F1_macro=1 would mean we archived the perfect result for both classes, while F1_macro=0.5 would mean that we failed on the minority class[35].

The implementation of 5 folds cross-validation for each of 59 candidates totaled 295 fits, which took approximately 6 days for a 2.4 GHz 32 GB RAM machine to process. The results, however, were not quite impressive. Increasing the maximum tree depth and the L2 regularization term generally improves the model both for 100 and 500 estimators. 500 estimators perform sufficiently better than 100 until the maximum tree depth does not exceed 10. If we set the maximum tree depth to 15, 100 estimators show higher results. Unfortunately, all the F1_marco scores vary only from 0,46 to 0,54, which means the minority class is almost suppressed by the majority. The best hyperparameters are {n_estimators=100, max_depth= 15, reg_lambda=10}, with a score of 0.54. The best score for {n_estimators=500, max_depth= 7 and reg_lambda=10} is quite close and equals to 0.531. The general idea is that we should use either lots of base learners or make each learner deeper. As the number of trees or their depth increases applying the L2- regularization term is needed to prevent overfitting. It should be also noted that the difference in score is minor compared to the duration of a 5-fold run: 80 mins for 500 estimators vs 41 mins for 100 estimators.

Consequently, the grid search for the ChP31" was enlightened to the randomized grid search with minimum parameters to explore (**Table 6**).

**Table 6.** The randomized grid search options for "ChP31".

| Selected Model Hyperparemeters | Grid | Description |
|---|---|---|
| n_estimators | 100 (default 100) | Number of boosted trees to fit. |
| max_depth | 7,10,15 (default 3) | Maximum tree depth for base learners |
| reg_lambda | 5,7,10 (default 1) | L2 regularization term on weights |
| scale_pos_weight | balance =330 (default 1) | Balancing of positive and negative weights |
| learning_rate | 0.1,0.5 (default 0.1) | Boosting learning rate |
| reg_alfa | 0 (default 0) | L1 regularization term on weights |

| Selected Model Hyperparemeters | Grid | Description |
|---|---|---|
| subsample | 0.8,1 (default 1) | Subsample ratio of the training instance |
| seed /random_state | 42 | Random number seed. Fixing the seed ensures we can reproduce the run |
| n_jobs/ nthread | -1 | The number of parallel threads used to run XGBoost. "-1"means using all possible threads. |

The best score was earned by the combination of {n_estimators=100, max_depth= 15, reg_lambda=10, subsample=1, learning rate=0.1} but, again, is quite low due to even more imbalanced dataset: F1_marco= 0.525.

The grid-search routines were not executed for the other models. They are quite time-consuming but the overall benefit in quality varies lightly. It might be assumed that the best (or close to them) combinations of hyperparameters would earn equally good results even if the dataset is a bit different. The combination applied to the models ChP21-29 was chosen rather empirically but appeared to be fairly close to the optimal: {n_estimators=500, max_depth= 7, reg_lambda=3, subsample=1, learning rate=0.1, scale_pos_weight=balance}.

### 7.6 ChangePoints performance overview

Below are the AUC and F-score results of the selected ChangePoints models ( Table 7).

Table 7. ChangePoints performance[3].

| Model Id | Train set | | | Test set | | | Balance |
|---|---|---|---|---|---|---|---|
| | AUC | F-score | # of Records | AUC | F-score | # of Records | |
| ChP20 | 100.00% | 100% (62%)* | 981 622 | 50.13% | 99% (0%)* | 420 405 | 357:1 |
| ChP22 | 96.30% | 98% (38%)* | 996 870 | 54.38% | 97 (3%)* | 427 019 | 76:1 |
| ChP26 | 96.23% | 93% (17%) | 1 170 211 | 84.53% | 92% (12%) | 505 377 | 78:1 |
| ChP27 | 97.75% | 94% (17%) | 994 288 | 84.54% | 94% (12%) | 426 623 | 98:1 |
| ChP28 | 99.96% | 98% (19%) | 982 401 | 88.62% | 98% (7%) | 420 487 | 290:1 |

---

[3] As part of an experiment the calculation for the model ChP20 was carried with a threshold level of 52% to decide between classes. The threshold of 60% was used for the model ChP22. This does not influence AUC figures, but slightly changes F-score (marked with "*"). For all the other models the default threshold of 50% was used.

| Model Id | Train set | | | Test set | | | Balance |
|---|---|---|---|---|---|---|---|
| | AUC | F-score | # of Records | AUC | F-score | # of Records | |
| ChP29 | 98.79% | 95% (14%) | 1 163 688 | 86.07% | 95% (8%) | 498 845 | 154:1 |
| ChP30 | 99.73% | 98% (32%) | 1 163 688 | 85.21% | 97% (10%) | 498 845 | 154:1 |
| ChP31 | 100% | 99% (34%) | 1 156 251 | 86,43% | 99% (6%) | 494 594 | 330:1 |

It should be noted that all the models, except the first two, show quite high AUC values on test sets. The low AUC scores (around 50%) mean that the predictions are almost random, probably due to the lack of trigger correction. The overall F-score is also very high, but the situation changes if we split the results between the majority and the minority classes. F1-score is the harmonic mean of precision and recall. Drilling down we can see that the low values of F1-score on the minority subset are mainly due to the extremely low precision - the rate of true positives among all predicted positives. This means, our model creates too many "false alarms", detecting non-existent change-points even on the train dataset. On the test set the recall value falls as well, meaning that a vast amount of true changepoints is also missed.

To illustrate this, we bring up the detailed characteristics of the model ChP29 (see **Table 8**), which demonstrated the best result on the pipeline. The results of the other models are quite similar.

**Table 8.** The detailed characteristics of the "ChP29" performance.

| | Train | | | Test | | |
|---|---|---|---|---|---|---|
| | Precision | Recall | F1-score | Precision | Recall | F1-score |
| Majority (New Trigger =0) | 100% | 92% | 96% | 100% | 92% | 95% |
| Minority (New Trigger =1) | **8%** | 99% | **14%** | **5%** | **58%** | **8%** |
| Avg/total | 99% | 92% | 95% | 99% | 91% | 95% |

On the other hand, we should remember, that AUC and F1-score are only the rough indicators of the ChangePoints model performance and we can make conclusions only after we run the pipeline.

## 8 TrendOrFlat model

Again, we shall start with the choice of features. We aim to recognize the tendencies inside the windows. The model should be suitable for any stock, market, time and –

ideally- window length. More than that, it should be capable of recognizing the tendency from its small patch. Below is the list of suggested features (5+1 target) and the intuition behind them (see **Table 9**).

Table 9. TrendOrFlat features

| Feature | Description |
| --- | --- |
| RegClose | The slope of the linear regression line for daily closing prices (logarithmic or not). The RegClose value for flat periods should be around zero and positive or negative for upward or downward tendencies respectively. The choice of daily closing prices (not open, high, or low) is backed by the experts' opinion: the closing price summarizes the events the day. |
| CloseR2 | The R2 coefficient of the linear regression line for daily closing prices (logarithmic or not). Indicates the degree of confidence in our regression estimates. If the slope is positive or negative but R2 is close to zero, then we should be less confident in the presence of a trend. |
| RegVol | The slope of the linear regression line for daily volumes (logarithmic or not). It is thought that volumes "support" trends. They rise in case of trend and stay stable if nothing happens. |
| VolR2 | The R2 coefficient of the linear regression line for daily volumes (logarithmic or not). Similarly to CloseR2, indicates the degree of confidence in our regression estimates. |
| LenTrend | The length of tendency (or the width of the window) in business days. We should be more confident in linear regression indicators if they are obtained from bigger datasets (longer tendencies) than for those which count several elements only. |
| NewTypeBool | Target variable. The Boolean analog of the "Type" field. It takes the value "1" in case of trend and "0" otherwise. |

These variables do not contain absolute values, which makes them suitable for any asset.

### 8.1 Dataset overview

The dataset for the TrendOrFlat model is different from that for the ChangePoints since now we are dealing with time periods, not separate trading days. The size of the dataset equals the total number of windows marked by experts. Since we opted for the train/test split by date, the split date is October 14th, 2014 for the models exploiting both Source I and Source II, and November 6th, 2014 if only Source II is used. We do not care about the contradictions this time, because we can hardly expect coinciding feature vectors – each window is somehow different. Even trigger correction is unnecessary because minor errors, while important for detecting the precise changepoints, will have minimal impact on the TrendOrFlat features. Neither do we care about the imbalance problem, since, as can be seen from the table, the ratio between positive and negative classes is roughly equal.

But what we should take into attention, is the ability of the model to recognize tendencies by their parts. To ensure this we supplemented the initial dataset, which contained full windows only, by feature vectors extracted from 5, 10, 20….90% parts of full windows. For example, to find the characteristics for the 20% part of the window, we simply take the first 20% of the tendency (in days) and calculate the feature values as if it was a full tendency.

This approach however might result in the appearance of super-short tendencies which can be considered as noise. To deal with this issue we drop off all the records with LenTrend<6. This choice is explained by the 5-day lag of the ChangePoints model, so anyway we won't need to recognize the tendencies that last less than 6 days. The characteristics of data sources for different TF models are given in Table 10.

**Table 10.** TrendOrFlat models with different dataset options[4].

| Model Id | Logarithmic | Source (split by date) | Experts | Balance | Number of full windows | Total number of records |
|---|---|---|---|---|---|---|
| TF9 | False | Source II | All | 1:1 | 16 759 | 166 912 |
| TF11 | True | Source I+II | All | 1,19:1 | 23 596 | 232 870 |
| TF12 | True | Source II | All | 1,02:1 | 16 116* | 162 224* |
| TF13 | True | Source I+II | D and G | 1,22:1 | 9 603 | 97 225 |

### 8.2 TrendOrFlat performance overview

TrendOrFlat model is also a binary classification model. We used the XGBoost algorithm again with the following set of hyperparameters: {n_estimators=100, max_depth=5, reg_lambda=3, learning_rate=0.2, seed=42, nthread=-1}. All the other hyperparameters were set to the algorithm default values. As we have noticed before, it is mainly the first three elements that influence the performance of the model. The number of base learners (n_estimators=100) and their maximum depth (max_depth=5) determine the level of complexity of the model while the L2-regularization term (reg_lambda=3) prevents overfitting if the model is too complex. From the ChangePoints model we know that even for a dataset containing a million records and 22 features, 100 estimators are enough. The TrendOrFlat dataset is much smaller, so no deep trees are required, consequently, the L2-regularization term can be set to a smaller value. The selected combination of hyperparameters provided good performance, so no grid search routines were executed (see **Table 11**).

---

[4] Due to technical reasons the period from October 14th to November 6th 2014 was excluded from the trainset of the TF12 model (marked with "*"). Both models TF11 and TF12 had October 14th 2014 as a breakpoint between train and test set. This explains the slight difference between the number of records in TF9 and TF12 datasets, which are based on the same source.

Table 11. TrendOrFlat performance overview.

| Model Id | Train set | | | | Test set | | | |
| --- | --- | --- | --- | --- | --- | --- | --- | --- |
| | AUC | F-score | Accuracy: | # of Records | AUC | F-score | Accuracy: | # of Records |
| TF9 | 83.41% | 75% | 75.29% | 166 912 | 76.25% | 69% | 69.65% | 101 215 |
| TF11 | 83.34% | 76% | 75.97% | 232 870 | 77.94% | 72% | 72.11% | 135 145 |
| TF12 | 83.44% | 75% | 75.23% | 162 224 | 76.17% | 70% | 70.42% | 102 112 |
| TF13 | 85.86% | 77% | 77.48% | 97 225 | 81.86% | 74% | 74.48% | 53 224 |

The F-score, precision and recall values do not vary within classes because the datasets are balanced, and all these metrics stay around 70-75% (**Table 11**). But we will see some difference if we look at the model performance for several tendency proportions. Below are the statistics for the TF13 model, the others behave similarly (**Table 12**). Even for very small parts of trends the classification quality is around 70% and sufficiently increases up to 90-95% when 80% days or more are shown to the model. That means that the TrendOrFlat model is likely to correct ChangePoints pitfalls. Also note, that if the output of the model is 1("Trend") we can easily determine the trend direction from the RegClose value (positive for upward trends and negative for downward).

Table 12. The behavior of the TF13 model for different tendency proportions.

| Proportion of tendency | Train set | | Test set | |
| --- | --- | --- | --- | --- |
| | AUC | Accuracy | AUC | Accuracy |
| 5.0% | 71.4% | 59.4% | 57.7% | 53.1% |
| 10.0% | 72.2% | 63.6% | 66.6% | 62.5% |
| 20.0% | 76.1% | 69.9% | 70.5% | 66.7% |
| 30.0% | 79.6% | 72.6% | 73.6% | 68.4% |
| 40.0% | 82.9% | 75.5% | 77.2% | 71.6% |
| 50.0% | 85.7% | 78.0% | 81.4% | 74.7% |
| 60.0% | 88.3% | 80.0% | 85.1% | 77.6% |
| 70.0% | 91.0% | 83.0% | 87.7% | 80.4% |
| 80.0% | 93.4% | 85.5% | 89.8% | 82.5% |
| 90.0% | 94.9% | 87.3% | 91.8% | 84.2% |
| 100.0% | 95.9% | 87.9% | 93.7% | 85.5% |

## 9    Pipeline results

The pipeline logic is described at the beginning of the paper. Here we shall concentrate on the discussion of the specific quality metrics.

Each time the upward trend is detected, we can virtually buy a unit of stock and sell it at the end of the tendency. Our profit, in percentage, can be calculated the following way:

$$Profit_{up} = \frac{Close_{t1} - Close_{t0}}{Close_{t0}},$$

where $Close_{t0}$ – is the daily closing price at the beginning of the trend, $Close_{t1}$ - the daily closing price at the end of the trend.– is the day closing price at the beginning of the trend, $Close_{t1}$ - the day closing price at the end of the trend.

For the downward trend, the formula is inverse: we first sell and then buy back for a lower price:

$$Profit_{up} = \frac{Close_{t0} - Close_{t1}}{Close_{t0}}.$$

Recalling, that the direction of the trend can be determined by the slope of the regression line, we can combine the two formulas:

$$Profit = sign(RegClose) \cdot \frac{Close_{t1} - Close_{t0}}{Close_{t0}},$$

where $sign(RegClose)$ is the sign of the RegClose value, calculated for that trend.

For each stock involved in the pipeline run the following statistics were calculated (**Table 13**):

Table 13. The basic pipeline performance indicators

| Indicator | Description |
|---|---|
| Profit | The sum of all profits earned during the time in position, in %. We open a long position for the upward trend, and a short – for the downward trend. The sum of *Profit_lng* and *Profit_sht* |
| Days_in | The total number of business days in position (the total length of all the identified trends). The sum of *Days_in_lng* and *Days_in_sht* |
| Times_in | The number of times the position was opened (the number of trends identified). The sum of *Times_in_lng* and *Times_in_sht*. |
| Profit_lng | The sum of all profits earned during the time in a long position, in % |
| Days_in_lng | Total number of business days in long position (the total length of all the identified upward trends) |
| Times_in_lng | The number of times the long position was opened (the number of upward trends identified) |
| Profit_sht | The sum of all profits earned during the time in a short position, in % |
| Days_in_sht | Total number of business days in short position (the total length of all the identified downward trends) |
| Times_in_sht | The number of times the long position was opened (the number of downward trends identified) |

Using these statistics, we can compare the behavior of the concurrent models for the same stocks and time periods. But the results vary significantly from stock to stock, so we will need more general metrics to compare (**Table 14**).

**Table 14.** The additional pipeline performance indicators

| Indicator | Description |
|---|---|
| numStocks | The number of stocks in the pipeline, % |
| Profit | The sum of all profits earned during the time in position (for all the stocks) |
| Days_in | The total number of business days in position (for all stocks) |
| Times_in | The number of times the position was opened (for all stocks |
| DayProfit | The profit per one day in position, % : DayProfit= Profit/ Days_in |
| YearProfit | dayProfit scaled per annum, %: YearProfit= DayProfit*250, where 250 is the average number of business days in a year |
| YearProfit_avg | The average annual profit, including the days not in position, %: YearProfit_avg = Profit/number of datapoints*250 |

To compare models, the last two indicators - *YearProfit* and *YearProfit_avg* - are the most informative, because they are independent of the length of the time period and the number of stocks in the pipeline. From a business point of view, it might also be important to see how many times we opened the positions or what was the proportion between short or long for each stock, because it influences the additional costs of trading.

*YearProfit* shows how much profit, in percentage, one would gain on the money invested into assets if it was working during the whole year. But in reality, trends alternate with flat periods, so the money is not working all the time. *YearProfit_avg* shows profit including flat periods when no investments are made and of course will be smaller than *YearProfit*. To be more precise, we should, however, not consider flat periods as zero-profitable, since money can be put on overnight deposits. But for the purpose of this paper, we will treat the *YearProfit_avg* indicator just as a measure of effectiveness in identifying trading opportunities which results in bigger total profits for a chosen set of stocks and time period.

The main pitfall of *YearProfit* and *YearProfit_avg* metrics is that they are not normalized, so solely by their values we can't definitely say what is a good result. But we can compare our results with those, calculated on the same datasets labelled by real experts or even the virtual "average" expert. We should bear in mind, however, that experts were dealing with historical data, or "saw the future", so their profits are bound to be much and much higher.

Below are the metrics, calculated for the different experts on two test sets (Table 15).

**Table 15.** The performance of experts (the ideal "ground truth")

| Experts | Source I+II (test set) | | Source II (test set) | |
|---|---|---|---|---|
| | YearProfit (%) | YearProfit_avg (%) | YearProfit (%) | YearProfit_avg (%) |
| A | 57,0 | 30,2 | 54,4 | 28,7 |
| B | 45,7 | 28,4 | 45,1 | 27,9 |

| Experts | Source I+II (test set) | | Source II (test set) | |
|---|---|---|---|---|
| | YearProfit (%) | YearProfit_avg (%) | YearProfit (%) | YearProfit_avg (%) |
| C | 61,8 | 17,9 | - | - |
| D | **62,5** | **27,9** | **62,5** | **27,4** |
| E | 30,6 | 15,9 | - | - |
| F | 39,3 | 17,8 | - | - |
| G | **70,5** | **26,8** | **68,6** | **26,9** |
| H | 49,3 | 32,1 | 45,8 | 31,4 |
| I | 49,9 | 20,8 | 46,4 | 19,1 |
| "Average" | 31,1 | 14,1 | 31,1 | 15,4 |

As we can notice, the experts D and G show the best results, while the performance of the "Average" expert (a product of our struggle with contradictions, when experts' opinions were averaged) is comparingly poor. This actually explains, why models, based on D and G experts labels (ChP28-30), perform better than others, and why averaging in models ChP20 and ChP31 didn't much improve the modelling results (though excluded contradictions).

The results of pipelines are presented below (**Table 16**). It was mentioned previously, that our ChangePoints models produce two much "false alerts", indicating non-existent changepoints. To deal with this we can increase the classification thresholds (default 0.5) to 0.6 or even 0.8 values. A couple of such experiments was run, though, unfortunately, this approach didn't bring much improvement: as the model also produces false negatives, we started missing the true changepoints.

**Table 16.** The pipeline performance

| ID Pipeline | Expert | Averaging | Train source (split by date) | Test source (split by date) | | Year Profit | Year Profit_avg |
|---|---|---|---|---|---|---|---|
| ChP20_TF9 | All | True | Source II | Source II | 0.65 | 10.1 | 4.1 |
| ChP22_TF9 | All | False | Source II | Source II | 0.65 | 10.8 | 4.4 |
| ChP26_TF11 | All | False | Source I+II | Source I | 0.85 | 6.4 | 0.2 |
| ChP26_TF11 | All | False | Source I+II | Source II | 0.85 | 16.7 | 3.6 |
| ChP26_TF11 | All | False | Source I+II | Source I | 0.5 | 19.3 | 0.1 |
| ChP26_TF11 | All | False | Source I+II | Source II | 0.5 | 28.9 | 1.5 |
| ChP27_TF12 | All | False | Source II | Source I | 0.85 | -0.2 | 0 |
| ChP27_TF12 | All | False | Source II | Source II | 0.85 | 8.4 | 2.7 |
| ChP28_TF12 | All | False | Source II | Source I | 0.5 | 4.1 | 0.1 |
| ChP28_TF12 | G | N/A | Source II | Source II | 0.5 | 11.3 | 2.1 |

| ID Pipeline | Expert | Averaging | Train source (split by date) | Test source (split by date) |  | Year Profit | Year Profit_avg |
|---|---|---|---|---|---|---|---|
| ChP28_TF12 | G | N/A | Source II | Source I | 0.7 | -4.1 | -0.2 |
| ChP28_TF12 | G | N/A | Source II | Source II | 0.7 | 9.1 | 2.6 |
| **ChP29_TF13** | **D and G** | **False** | **Source I+II** | **Source I+II** | **0.5** | **28.8** | **6.9** |
| Ch30_TF13 | D and G | False | Source I+II | Source I+II | 0.5 | 22.6 | 6.1 |
| Ch31_TF13 | D and G | True | Source I+II | Source I+II | 0.5 | 18.1 | 5.8 |

Clearly, the best result is obtained by the combination of ChP29 and TF13 models (highlighted in bold). This pipeline shows both best *YearProfit* and *YearProfit_avg* values. Of course, they are much more modest than ground truth, but, again, the experts were using historical data. Nevertheless, having nearly 30% on investment per annum looks quite impressive. Unfortunately, the model still did miss a lot of opportunities (76% of records are predicted as flat), so *YearProfit_avg* is not too big. If we add overnight interest paid on flat periods (say, 7%) our average profit will reach 7%*0.76+6,9%=12,2% per annum. Illustrations of model performance are presented are presented in **Fig. 9** and **Fig. 10**.

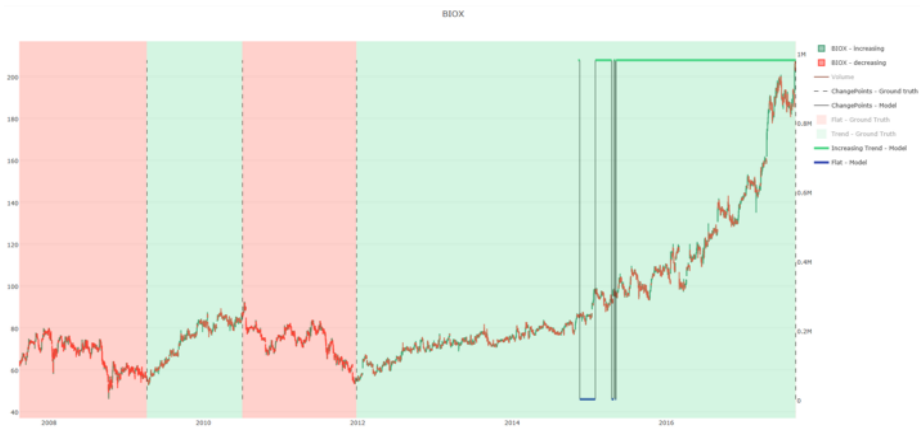

**Fig. 9.** An example of model performance (BIOX)

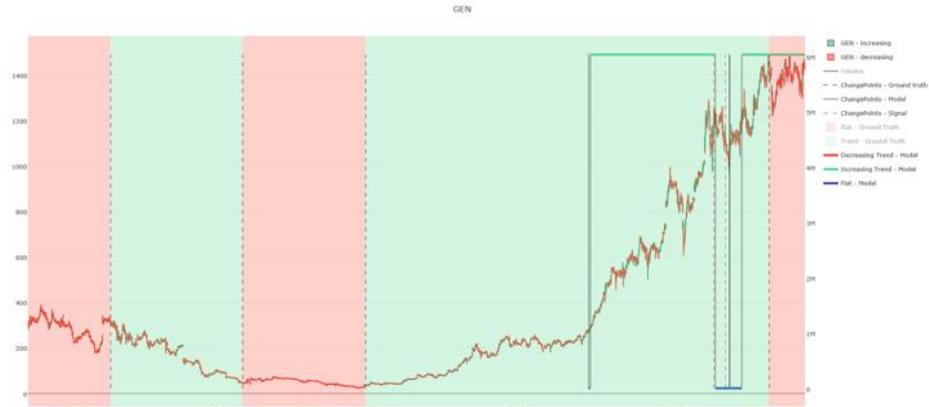

**Fig. 10.** An example of model performance (GEN)

## Conclusion

The model presented in the research can be used by both individual and institutional investors. It produces "buy" and "sell" signals when starting or endpoints of trends are identified. The profit earned on days in position can reach 28.8% per annum, but definitely, the result can be improved. Unlike in the traditional approaches, the labels (trend or flat) are not derived from prices but filled manually by experts who worked with stock data as with images. The task is quite challenging since we actually try to 'digitalize' successful traders' skills and we can only compare the performance of the model with the performance of the experts themselves. The comparison with the results of other researchers would be inadequate since we use a completely different source for ground truth.

There are several directions for this work:

─ Implementing other approaches to deal with contradicting labels and the imbalanced datasets – two major issues which influence the quality of the ChangePoints model. Though the XGboost algorithm tries to correct the proportion between classes, obviously this is not enough. Perhaps certain undersampling technics might improve the situation.
─ Selecting other sets of features for the ChangePoints and TrendOrFlat models. Various combinations of technical indicators should be tried for ChangePoints models and probably different time lags and threshold levels. As for TrendOrFlat, the improvement should be concentrated on <u>early</u> tendency identification, i.e. recognizing trends shortly after their start.
─ Experimenting with trading strategies, for example, excluding short positions as less profitable and more complicated, ignoring flat periods - that is, closing positions only when we have strong evidence of a new trend, etc.

− And finally, totally changing the model structure and using other machine learning algorithms, for example, convolutional neural networks, can also sufficiently improve the model.

**References**

1. Henrique, B.M., Sobreiro, V.A., Kimura, H.: Literature review: Machine learning techniques applied to financial market prediction. Expert Syst. Appl. (2019). https://doi.org/10.1016/j.eswa.2019.01.012
2. Brink, H., and Richards, J.: Real-World Machine Learning Version 4. (2014)
3. Thakkar, A., Chaudhari, K.: Fusion in stock market prediction: A decade survey on the necessity, recent developments, and potential future directions. Inf. Fusion. 65, (2021). https://doi.org/10.1016/j.inffus.2020.08.019
4. Nair, B.B., Dharini, N.M., Mohandas, V.P.: A stock market trend prediction system using a hybrid decision tree-neuro-fuzzy system. In: Proceedings - 2nd International Conference on Advances in Recent Technologies in Communication and Computing, ARTCom 2010 (2010)
5. Nair, B.B., Mohandas, V.., Sakthivel, N.R.: A Decision tree- Rough set Hybrid System for Stock Market Trend Prediction. Int. J. Comput. Appl. (2010). https://doi.org/10.5120/1106-1449
6. Nair, B.B., Sakthivel, V.P.M.N.R.: A Genetic Algorithm Optimized Decision Tree- SVM based Stock Market Trend Prediction System. Int. J. (2010)
7. Paliyawan, P.: Stock market direction prediction using data mining classification. ARPN J. Eng. Appl. Sci. (2015)
8. Patel, J., Shah, S., Thakkar, P., Kotecha, K.: Predicting stock and stock price index movement using Trend Deterministic Data Preparation and machine learning techniques. Expert Syst. Appl. (2015). https://doi.org/10.1016/j.eswa.2014.07.040
9. Qin, Q., Wang, Q.-G., Li, J., Ge, S.S.: Linear and Nonlinear Trading Models with Gradient Boosted Random Forests and Application to Singapore Stock Market. J. Intell. Learn. Syst. Appl. (2013). https://doi.org/10.4236/jilsa.2013.51001
10. Ismail, M.S., Md Noorani, M.S., Ismail, M., Abdul Razak, F., Alias, M.A.: Predicting next day direction of stock price movement using machine learning methods with persistent homology: Evidence from Kuala Lumpur Stock Exchange. Appl. Soft Comput. J. (2020). https://doi.org/10.1016/j.asoc.2020.106422
11. Ma, Y., Han, R., Wang, W.: Portfolio optimization with return prediction using deep learning and machine learning. Expert Syst. Appl. (2021). https://doi.org/10.1016/j.eswa.2020.113973
12. Bisoi, R., Dash, P.K., Parida, A.K.: Hybrid Variational Mode Decomposition and evolutionary robust kernel extreme learning machine for stock price and movement prediction on daily basis. Appl. Soft Comput. J. 74, (2019). https://doi.org/10.1016/j.asoc.2018.11.008

13. Zhou, F., Zhang, Q., Sornette, D., Jiang, L.: Cascading logistic regression onto gradient boosted decision trees for forecasting and trading stock indices. Appl. Soft Comput. J. (2019). https://doi.org/10.1016/j.asoc.2019.105747
14. Yang, J., Zhao, C., Yu, H., Chen, H.: Use GBDT to Predict the Stock Market. Procedia Comput. Sci. 174, (2020). https://doi.org/10.1016/j.procs.2020.06.071
15. Soujanya, R., Akshith Goud, P., Bhandwalkar, A., Anil Kumar, G.: Evaluating future stock value asset using machine learning. In: Materials Today: Proceedings (2020)
16. Lee, T.K., Cho, J.H., Kwon, D.S., Sohn, S.Y.: Global stock market investment strategies based on financial network indicators using machine learning techniques. Expert Syst. Appl. 117, (2019). https://doi.org/10.1016/j.eswa.2018.09.005
17. Paiva, F.D., Cardoso, R.T.N., Hanaoka, G.P., Duarte, W.M.: Decision-making for financial trading: A fusion approach of machine learning and portfolio selection. Expert Syst. Appl. 115, (2019). https://doi.org/10.1016/j.eswa.2018.08.003
18. Yang, S.Y., Yu, Y., Almahdi, S.: An investor sentiment reward-based trading system using Gaussian inverse reinforcement learning algorithm. Expert Syst. Appl. 114, (2018). https://doi.org/10.1016/j.eswa.2018.07.056
19. Chatzis, S.P., Siakoulis, V., Petropoulos, A., Stavroulakis, E., Vlachogiannakis, N.: Forecasting stock market crisis events using deep and statistical machine learning techniques. Expert Syst. Appl. 112, (2018). https://doi.org/10.1016/j.eswa.2018.06.032
20. Long, J., Chen, Z., He, W., Wu, T., Ren, J.: An integrated framework of deep learning and knowledge graph for prediction of stock price trend: An application in Chinese stock exchange market. Appl. Soft Comput. J. 91, (2020). https://doi.org/10.1016/j.asoc.2020.106205
21. Moews, B., Ibikunle, G.: Predictive intraday correlations in stable and volatile market environments: Evidence from deep learning. Phys. A Stat. Mech. its Appl. 547, (2020). https://doi.org/10.1016/j.physa.2020.124392
22. Vijh, M., Chandola, D., Tikkiwal, V.A., Kumar, A.: Stock Closing Price Prediction using Machine Learning Techniques. In: Procedia Computer Science (2020)
23. Zhou, F., Zhou, H. min, Yang, Z., Yang, L.: EMD2FNN: A strategy combining empirical mode decomposition and factorization machine based neural network for stock market trend prediction. Expert Syst. Appl. 115, (2019). https://doi.org/10.1016/j.eswa.2018.07.065
24. Hoseinzade, E., Haratizadeh, S.: CNNpred: CNN-based stock market prediction using a diverse set of variables. Expert Syst. Appl. 129, (2019). https://doi.org/10.1016/j.eswa.2019.03.029
25. Picasso, A., Merello, S., Ma, Y., Oneto, L., Cambria, E.: Technical analysis and sentiment embeddings for market trend prediction. Expert Syst. Appl. 135, (2019). https://doi.org/10.1016/j.eswa.2019.06.014
26. Chandrinos, S.K., Sakkas, G., Lagaros, N.D.: AIRMS: A risk management tool

using machine learning. Expert Syst. Appl. 105, (2018). https://doi.org/10.1016/j.eswa.2018.03.044
27. Ma, Y., Han, R., Wang, W.: Portfolio optimization with return prediction using deep learning and machine learning. Expert Syst. Appl. 165, (2021). https://doi.org/10.1016/j.eswa.2020.113973
28. Soujanya, R., Goud, P.A., Bhandwalkar, A., Kumar, G.A.: Evaluating future stock value asset using machine learning. Mater. Today Proc. (2020). https://doi.org/10.1016/j.matpr.2020.08.385
29. Ismail, M.S., Md Noorani, M.S., Ismail, M., Abdul Razak, F., Alias, M.A.: Predicting next day direction of stock price movement using machine learning methods with persistent homology: Evidence from Kuala Lumpur Stock Exchange. Appl. Soft Comput. J. 93, (2020). https://doi.org/10.1016/j.asoc.2020.106422
30. Weng, B., Lu, L., Wang, X., Megahed, F.M., Martinez, W.: Predicting short-term stock prices using ensemble methods and online data sources. Expert Syst. Appl. 112, (2018). https://doi.org/10.1016/j.eswa.2018.06.016
31. Jiang, M., Liu, J., Zhang, L., Liu, C.: An improved Stacking framework for stock index prediction by leveraging tree-based ensemble models and deep learning algorithms. Phys. A Stat. Mech. its Appl. 541, (2020). https://doi.org/10.1016/j.physa.2019.122272
32. Chen, T., Guestrin, C.: XGBoost: A scalable tree boosting system. In: Proceedings of the ACM SIGKDD International Conference on Knowledge Discovery and Data Mining (2016)
33. XGBoost on GitHub Repository, https://github.com/dmlc/xgboost/tree/master/demo#machine-learning-challenge-winning-solutions
34. XGBoost Parameters, https://xgboost.readthedocs.io/en/latest/parameter.html
35. Scikit-learn Metrics and Scoring, https://scikit-learn.org/stable/modules/model_evaluation.html